# An Almost Unbiased Estimator for Population Mean using Known Value of Population Parameter(s)

*Rajesh Singh*[1], *S.B. Gupta*[2] *and Sachin Malik*[2, *]

[1] Department of Statistics, Banaras Hindu University, Varanasi-221005, India

[2]Department of Community Medicine, SRMS Institute of Medical Sciences, Bareilly- 243202, India

**Abstract:** In this paper we have proposed an almost unbiased estimator using known value of some population parameter(s) with known population proportion of an auxiliary variable. A class of estimators is defined which includes [1], [2] and [3] estimators. Under simple random sampling without replacement (SRSWOR) scheme the expressions for bias and mean square error (MSE) are derived. Numerical illustrations are given in support of the present study.

**Key words**: Auxiliary information, proportion, bias, mean square error, unbiased estimator.

## 1. Introduction

It is well known that the precision of the estimates of the population mean or total of the study variable y can be considering improved by the use of known information on an auxiliary variable x which is highly correlated with the study variable y. Out of many methods ratio, product and regression methods of estimation are good illustrations in this context. Using known values of certain population's parameters several authors have proposed improved estimators including [4, 5, 6, 7, 8, 9, 10, 11, 12, 13].

In many practical situations, instead of existence of auxiliary variables there exit some

* Corresponding author e-mail: sachinkurava999@gmail.com

auxiliary attributes $\phi$ (say), which are highly correlated with the study variable y, such as

i. Amount of milk produced (y) and a particular breed of cow ($\phi$).

ii. Sex ($\phi$) and height of persons (y) and

iii. Amount of yield of wheat crop and a particular variety of wheat ($\phi$) etc. (see [14]).

Many more situations can be encountered in practice where the information of the population mean $\overline{Y}$ of the study variable y in the presence of auxiliary attributes assumes importance. For these reasons various authors such as [15, 16, 17, 18, 19] have paid their attention towards the improved estimation of population mean $\overline{Y}$ of the study variable y taking into consideration the point biserial correlation between a variable and an attribute.

Let $A = \sum_{i=1}^{N} \phi_i$ and $a = \sum_{i=1}^{n} \varphi_i$ denote the total number of units in the population and sample possessing attribute $\phi$ respectively, $P = \frac{A}{N}$ and $p = \frac{a}{n}$ denote the proportion of units in the population and sample, respectively, possessing attribute $\phi$.

To estimate $\overline{Y}$, the usual estimator is given by

$$\mathrm{Var}(\bar{y}) = f_1 S_y^2 \tag{1}$$

Define,

$$e_y = \frac{(\bar{y} - \overline{Y})}{\overline{Y}} \quad e_\phi = \frac{(p - P)}{P},$$

$$E(e_i) = 0, (i = y, \phi)$$

$$E(e_y^2) = f_1 C_y^2, \; E(e_\varphi^2) = f_1 C_p^2,$$

$$E(e_y e_\varphi) = f_1 \rho_{pb} C_y C_{p.}$$

Where,

$$f_1 = \left(\frac{1}{n} - \frac{1}{N}\right),\ C_y^2 = \frac{S_y^2}{\bar{Y}^2}, \qquad C_p^2 = \frac{S_p^2}{P^2},$$

and $\rho_{pb} = \frac{S_{y\phi}}{S_y S_\phi}$ is the point bi-serial correlation coefficient.

Here,

$$S_y^2 = \frac{1}{N-1}\sum_{i=1}^{N}\left(y_i - \bar{Y}\right)^2,\quad S_\phi^2 = \frac{1}{N-1}\sum_{i=1}^{N}(\phi_i - P)^2$$

and $S_{y\phi} = \frac{1}{N-1}\left(\sum_{i=1}^{N} y_i\phi_i - NP\bar{Y}\right), f_2 = \left(\frac{1}{n'} - \frac{1}{N}\right),$

In order to have an estimate of the study variable y, assuming the knowledge of the population proportion P, [1] proposed the following estimator

$$t_{NGR} = \bar{y}\left(\frac{P}{p}\right) \tag{2}$$

$$t_{NGP} = \bar{y}\left(\frac{p}{P}\right) \tag{3}$$

Following [1] , we propose the following estimator

$$t_{S1} = \bar{y}\left(\frac{K_1P + K_2K_3}{K_1p + K_2K_3}\right)^{\alpha} \tag{4}$$

The Bias and MSE expressions of the estimator $t_{S1}$ up to the first order of approximation are, respectively, given by

$$B(t_{S1}) = \bar{Y} f_1 C_p^2 \left[ \frac{\alpha(\alpha+1)V_1^2}{2} - \alpha V_1 K_p \right] \tag{5}$$

$$MSE(t_{S1}) = \bar{Y}^2 f_1 \left[ C_y^2 + C_p^2 \left( \alpha^2 V_1^2 - 2V_1 \alpha K_{p_1} \right) \right] \tag{6}$$

Also following [2], we propose the following estimator

$$t_{S2} = \bar{y} \left\{ 2 - \left( \frac{p}{P} \right)^{\beta} \exp\left[ \lambda \left( \frac{(K_4 P + K_5) - (K_4 p + K_5)}{(K_4 P + K_5) + (K_4 p + K_5)} \right) \right] \right\} \tag{7}$$

The Bias and MSE expressions of the estimator $t_{S2}$ up to the first order of approximation are, respectively, given by

$$B(t_{S2}) = \bar{Y} f_1 C_p^2 \left[ \frac{\lambda V_2 \beta}{2} - \frac{\beta(\beta-1)}{2} - \frac{\lambda(\lambda+2)V_2^2}{8} - \beta K_p + \frac{\lambda V_2 K_p}{2} \right] \tag{8}$$

$$MSE(t_{S2}) = \bar{Y}^2 f_1 \left[ C_y^2 + C_p^2 \left( \beta^2 + \frac{\lambda^2 V_2^2}{4} - \beta \lambda V_2 \right) - 2K_p C_p^2 \left( \beta - \frac{\lambda V_2}{2} \right) \right] \tag{9}$$

$\alpha$, $\lambda$ and $\beta$ are suitable chosen constants. Also $K_1, K_3, K_4, K_5$ are either real numbers or function of known parameters of the auxiliary attributes $\phi$ such as $C_p$, $\beta_2(\phi)$, $\rho_{pb}$ and $K_P$. $K_2$ is an integer which takes values +1 and -1 for designing the estimators and

$$\left. \begin{aligned} V_1 &= \frac{K_1 P}{K_1 P + K_2 K_3} \\ V_2 &= \frac{K_4 P}{K_4 P + K_5} \end{aligned} \right\}$$

We see that the estimator's $t_{S1}$ and $t_{S2}$ are biased estimators. In some applications bias is disadvantageous. Following these estimators we have proposed almost unbiased estimator of $\bar{Y}$.

## 2. Almost unbiased estimator

Suppose, $t_{S0}=\bar{y}, t_{S1}=\bar{y}\left(\frac{K_1P+K_2K_3}{K_1p+K_2K_3}\right)^{\alpha}, t_{S2}=\bar{y}\left\{2-\left(\frac{p}{P}\right)^{\beta}\exp\left[\lambda\left(\frac{(K_4P+K_5)-(K_4p+K_5)}{(K_4P+K_5)+(K_4p+K_5)}\right)\right]\right\}$

Such that $t_{S0}, t_{S1}, t_{S2}\in W$, where W denotes the set of all possible estimators for estimating the population mean $\bar{Y}$. By definition, the set W is a linear variety if

$$t_p=\sum_{i=0}^{2}w_i t_{Si}\in W \tag{10}$$

Such that, $\sum_{i=0}^{2}w_i=1$ and $w_i\in R$ (11)

where, $w_i\,(i=0,1,2)$ denotes the constants used for reducing the bias in the class of estimators, W denotes the set of those estimators that can be constructed from $t_{Si}\,(i=0,1,2)$ and R denotes the set of real numbers.

Expressing equation (10) in terms of e's, we have

$$t_p=\bar{Y}\left[1+e_y+w_1\left(\frac{\alpha(\alpha+1)V_1^2e_\phi^2}{2}-\alpha V_1e_\phi-\alpha V_1e_ye_\phi\right)\right.$$

$$\left.+w_2\left(-\beta e_\phi-\frac{\beta(\beta-1)e_\phi^2}{2}+\frac{\lambda V_2e_\phi}{2}+\frac{\lambda V_2\beta e_\phi^2}{2}-\frac{\lambda(\lambda+2)V_2^2e_\phi^2}{8}-\beta e_ye_\phi+\frac{\lambda V_2e_ye_\phi}{2}\right)\right] \tag{12}$$

Subtracting $\bar{Y}$ from both sides of equation (12) and then taking expectation of both sides, we get the bias of the estimator $t_p$ up to the first order of approximation, as

$$B(t_p) = \bar{Y} f_1 w_1 C_p^2 \left( \frac{\alpha(\alpha+1)V_1^2}{2} - \alpha V_1 K_p \right) + \bar{Y} f_1 w_2 C_p^2 \left( \frac{\lambda V_2 \beta}{2} - \frac{\beta(\beta-1)}{2} - \frac{\lambda(\lambda+2)V_2^2}{8} - \beta K_P + \frac{\lambda V_2 K_P}{2} \right) \quad (13)$$

From (12), we have

$$(t_p - \bar{Y}) = \bar{Y}\left[ e_y - w_1 \alpha V_1 e_\varphi - w_2 \left( \beta e_\varphi + \frac{\lambda V_2 e_\varphi}{2} \right) \right] \quad (14)$$

Squaring both sides of (14) and then taking expectation, we get the MSE of the estimator $t_p$ up to the first order of approximation, as

$$MSE(t_p) = \bar{Y}^2 f_1 \left[ C_y^2 + C_p^2 \left( Q^2 - 2QK_p \right) \right] \quad (15)$$

Which is minimum when

$$Q = K_p \quad (16)$$

where, $Q = w_1 \alpha V_1 + w_2 \left( \beta - \frac{\lambda V_2}{2} \right)$ (17)

Putting the value of $Q = K_p$ in (15), we have optimum value of estimator as $t_p$ (optimum).

Thus the minimum MSE of $t_p$ is given by

$$\min . MSE(t_p) = \bar{Y}^2 f_1 C_y^2 \left( 1 - \rho_{pb}^2 \right) \quad (18)$$

Which is same as that of traditional linear regression estimator.

from (11) and (17), we have only two equations in three unknowns. It is not possible to find the unique values for $w_i$'s, i=0, 1, 2. In order to get unique values of $w_i$'s, we shall impose the linear restriction

$$\sum_{i=0}^{2} w_i B(t_{Si}) = 0 \tag{19}$$

where, $B(t_{Si})$ denotes the bias in the $i^{th}$ estimator.

Equations (11), (17) and (19) can be written in the matrix form as

$$\begin{bmatrix} 1 & 1 & 1 \\ 0 & \alpha V_1 & \beta - \frac{\lambda V_2}{2} \\ 0 & B(t_{S1}) & B(t_{S2}) \end{bmatrix} \begin{bmatrix} w_0 \\ w_1 \\ w_2 \end{bmatrix} = \begin{bmatrix} 1 \\ k_p \\ 0 \end{bmatrix} \tag{20}$$

Using (20), we get the unique values of $w_i$'s, i=0, 1, 2 as

$$\left.\begin{aligned} w_0 &= \frac{\alpha V_1[\alpha V_1 A_2 - A_1 X_1] - X_1 K_P A_1 - X_2 \alpha V_1[\alpha V_1 A_2 - A_1 X_1] - \alpha V_1 K_P A_1}{\alpha V_1[\alpha V_1 A_2 - A_1 X_1]} \\ w_1 &= \frac{X_1 K_P A_1}{\alpha V_1[\alpha V_1 A_2 - A_1 X_1]} + X_2 \\ w_2 &= \frac{K_P A_1}{[\alpha V_1 A_2 - A_1 X_1]} \end{aligned}\right\}$$

$$\left.\begin{aligned} A_1 &= \frac{\alpha(\alpha+1)V_1^2}{2} - \alpha V_1 K_p \\ A_2 &= \frac{\lambda V_2 \beta}{2} - \frac{\beta(\beta-1)}{2} - \frac{\lambda(\lambda+2)V_2^2}{8} - \beta K_P + \frac{\lambda V_2 K_P}{2} \\ X_1 &= A_1\left[\beta - \frac{\lambda V_2}{2}\right] \\ X_2 &= \frac{K_p}{\alpha V_1} \end{aligned}\right\}$$

Use of these $w_i$'s, i=0, 1, 2 remove the bias up to terms of order $o(n^{-1})$ at (10).

## 3.Empirical study

For empirical study we use the data sets earlier used by [20] (population 1) and [21] (population 2) to verify the theoretical results.

**Data statistics:**

| **Population** | N | n | $\bar{Y}$ | P | $C_y$ | $C_p$ | $\rho_{pb}$ | $\beta_2(\phi)$ |
|---|---|---|---|---|---|---|---|---|
| **Population I** | 89 | 20 | 3.360 | 0.1236 | 0.60400 | 2.19012 | 0.766 | 6.2381 |
| **Population II** | 25 | 10 | 9.44 | 0.400 | 0.17028 | 1.27478 | -0.387 | 4.3275 |

**Table 3.1**: Values of $w_i$'s,

| $w_i$'s, | **Population 1** | **Population 2** |
|---|---|---|
| $w_0$ | -3.95624 | 1.124182 |
| $w_1$ | 5.356173 | 0.020794 |
| $w_2$ | -0.39993 | -0.14498 |

**Table 3.2: PRE of different estimators of $\overline{Y}$ with respect to $\bar{y}$**

| **Choice of scalars** | | | | | | | | | | | | | |
|---|---|---|---|---|---|---|---|---|---|---|---|---|---|
| $w_0$ | $w_1$ | $w_2$ | $K_1$ | $K_2$ | $K_3$ | $K_4$ | $K_5$ | $\alpha$ | $\beta$ | $\lambda$ | **Estimator** | **PRE (POPI)** | **PRE (POPII)** |
| 1 | 0 | 0 | | | | | | | | | $\bar{y}$ | 100 | 100 |
| 0 | 1 | 0 | 1 | 1 | 0 | | | 1 | | | $t_{NGR}$ | 11.63 | 1.59 |
| | | | 1 | 1 | 0 | | | -1 | | | $t_{NGP}$ | 5.075 | 1.94 |
| 0 | 0 | 1 | | | | | | | 1 | 0 | $t_{1(1,0)}$ | 12.88 | 1.59 |
| | | | | | | | | | -1 | 0 | $t_{1(-1,0)}$ | 5.43 | 1.95 |
| | | | | | | 1 | 0 | | 1 | 1 | $t_{2(1,1)}$ | 73.59 | 0.84 |
| | | | | | | 1 | 0 | | 1 | -1 | $t_{2(1,-1)}$ | 4.94 | 8.25 |
| | | | | | | 1 | 0 | | 0 | 1 | $t_{2(0,1)}$ | 14.95 | 8.25 |
| | | | | | | 1 | 0 | | 0 | -1 | $t_{2(0,-1)}$ | 73.48 | 5.58 |
| $w_0$ | $w_1$ | $w_2$ | 1 | 1 | 1 | 1 | 1 | 1 | 1 | 1 | $t_P$ **optimum** | 241.98 | 117.61 |

## 4. Proposed estimators in two phase sampling

In some practical situations when P is not known a priori, the technique of two-phase sampling is used. Let $p'$ denote the proportion of units possessing attribute $\phi$ in the first phase sample of size $n'$; $p$ denote the proportion of units possessing attribute $\phi$ in the second phase sample of size $n' > n$ and $\bar{y}$ denote the mean of the study variable y in the second phase sample.

In two-phase sampling the estimator $t_p$ will take the following form

$$t_{pd} = \sum_{i=0}^{2} h_i t_{id} \in H \tag{21}$$

$$\text{Such that, } \sum_{i=0}^{2} h_i = 1 \quad \text{and } h_i \in R \tag{22}$$

Where,

$$t_{0d} = \bar{y}, \; t_{1d} = \bar{y}\left(\frac{K_1 p' + K_2 K_3}{K_1 p + K_2 K_3}\right)^m, \; t_{2d} = \bar{y}\left\{2 - \left(\frac{p}{p'}\right)^q \exp\left[\gamma\left(\frac{(K_4 p' + K_5) - (K_4 p + K_5)}{(K_4 p' + K_5) + (K_4 p + K_5)}\right)\right]\right\}$$

The Bias and MSE expressions of the estimator $t_{1d}$ and $t_{2d}$ up to the first order of approximation are, respectively, given by

$$B(t_{1d}) = \bar{Y}\left[\frac{m(m-1)R_1^2 f_2 C_p^2}{2} + \frac{m(m+1)R_1^2 f_1 C_p^2}{2} - m^2 R_1^2 f_2 C_p^2 + m R_1 f_3 k_p C_p^2\right] \tag{23}$$

$$MSE(t_{1d}) = \bar{Y}^2\left[f_1 C_y^2 + m^2 R_1^2 f_3 C_p^2 - 2m R_1 k_p f_3 C_p^2\right] \tag{24}$$

$$B(t_{2d}) = \overline{Y}\left[-\frac{q(q-1)f_1C_p^2}{2} + \frac{q(q+1)f_2C_p^2}{2} + qf_2k_pC_p^2 + q^2f_2C_p^2 + f_3\gamma R_2k_pC_p^2 + f_3\gamma R_2qC_p^2\right] \quad (25)$$

$$MSE(t_{2d}) = \overline{Y}^2\left[f_1C_y^2 + L_1^2f_3C_p^2\right] \quad (26)$$

Where,

$$\left.\begin{aligned} R_1 &= \frac{K_1P}{K_1P + K_2K_3} \\ R_2 &= \frac{K_4P}{2[K_4P + K_5]} \\ L_1 &= q - \gamma A_2 \end{aligned}\right\} \quad (27)$$

Expressing (21) in terms of e's, we have

$$t_{pd} = \overline{Y}\left[1 + e_y + w_1\left(\frac{m(m+1)R_1^2e_\varphi^2}{2} - mR_1e_\varphi - mR_1e_ye_\varphi + mR_1e_ye_\varphi + \frac{m(m-1)R_1^2e'^2_\varphi}{2} + mR_1e_ye'_\varphi\right)\right.$$
$$\left. + w_2\left(-qe_\varphi - \frac{q(q-1)e_\varphi^2}{2} + qe'_\varphi + q^2e_\varphi e'_\varphi - \frac{q(q+1)e'^2_\varphi}{2} - \gamma R_2(e'_\varphi - e_\varphi) + \gamma R_2(e_ye_\varphi - e_ye'_\varphi) - qe_ye_\varphi\right)\right]$$

Subtracting $\overline{Y}$ from both sides of the above equation and then taking expectation of both sides, we get the bias of the estimator $t_{pd}$ up to the first order of approximation, as

$$B(t_{pd}) = \overline{Y}[B(t_{1d}) + B(t_{2d})] \quad (28)$$

Also,

$$(t_{pd} - \overline{Y}) = \overline{Y}\left[e_0 + w_1[mR_1e'_\varphi - mR_1e_\varphi] + w_2(-qe_\varphi + qe'_\varphi - \gamma R_2e'_\varphi + \gamma R_2e_\varphi)\right] \quad (29)$$

Squaring both sides of (29) and then taking expectation, we get the MSE of the estimator $t_{pd}$ up to the first order of approximation, as

$$MSE(t_{pd}) = \overline{Y}^2\left[f_1C_y^2 + L_2^2 f_3 C_p^2 - 2L_2 f_3 k_p C_p^2\right] \quad (30)$$

Which is minimum when

$$L_2 = K_p \quad (31)$$

Where $L_2 = h_1 mR_1 + h_2(q - \gamma R_2)$ (32)

Putting the value of $L_2 = K_p$ in (30), we have optimum value of estimator as $t_{pd}$ (optimum).

Thus the minimum MSE of $t_{pd}$ is given by

$$\min.MSE(t_{pd}) = \overline{Y}^2 C_y^2 \left(f_1 - f_3 \rho_{pb}^2\right) \quad (33)$$

which is same as that of traditional linear regression estimator.

from (22) and (32), we have only two equations in three unknowns. It is not possible to find the unique values for $h_i$'s, 1=0, 1, 2. In order to get unique values of $h_i$'s, we shall impose the linear restriction

$$\sum_{i=0}^{2} h_i B(t_{id}) = 0 \quad (34)$$

where, $B(t_{id})$ denotes the bias in the $i^{th}$ estimator.

Equations (22), (32) and (34) can be written in the matrix form as

$$\begin{bmatrix} 1 & 1 & 1 \\ 0 & mR_1 & q - \gamma R_2 \\ 0 & B(t_{1d}) & B(t_{2d}) \end{bmatrix} \begin{bmatrix} h_0 \\ h_1 \\ h_2 \end{bmatrix} = \begin{bmatrix} 1 \\ k_p \\ 0 \end{bmatrix} \quad (35)$$

Using (35), we get the unique values of $h_i$'s, i=0, 1, 2 as

$$\left.\begin{aligned} h_0 &= 1 - h_1 - h_2 \\ h_1 &= \frac{k_p}{mR_1} - \frac{N_1 K_P (q - \gamma R_2)}{N_1 q - mR_1 N_2 - N_1 \gamma R_2} \\ h_2 &= \frac{K_P N_1}{[N_1 q - mR_1 N_2 - N_1 \gamma R_2]} \end{aligned}\right\}$$

where,

$$\left.\begin{aligned} N_1 &= \frac{m(m-1)R_1^2 f_2 C_p^2}{2} + \frac{m(m+1)R_1^2 f_1 C_p^2}{2} - m^2 R_1^2 f_2 C_p^2 + mR_1 f_3 k_p C_p^2 \\ N_2 &= \left[ -\frac{q(q-1)f_1 C_p^2}{2} + \frac{q(q+1)f_2 C_p^2}{2} + qf_2 k_p C_p^2 + q^2 f_2 C_p^2 + f_3 \gamma R_2 k_p C_p^2 + f_3 \gamma R_2 q C_p^2 \right] \end{aligned}\right\}$$

Use of these $h_i$'s, i=0, 1, 2 remove the bias up to terms of order $o(n^{-1})$ at (21).

## 5. Empirical Study

For empirical study we use the data sets earlier used by [20] (population 1) and [21] (population 2) to verify the theoretical results.

**Data statistics:**

| **Pop.** | N | n | $\bar{Y}$ | P | $p'$ | $C_y$ | $C_p$ | $\rho_{pb}$ | $n'$ |
|---|---|---|---|---|---|---|---|---|---|
| **Pop.I** | 89 | 23 | 1322 | 0.1304 | 0.13336 | 0.69144 | 2.7005 | 0.408 | 45 |
| **Pop.II** | 25 | 7 | 7.143 | 0.294 | 0.308 | 0.36442 | 1.3470 | -0.314 | 13 |

**Table 5.1: PRE of different estimators of $\overline{Y}$ with respect to $\bar{y}$**

| **Choice of scalars** | | | | | | | | | | | | | |
|---|---|---|---|---|---|---|---|---|---|---|---|---|---|
| $w_0$ | $w_1$ | $w_2$ | $K_1$ | $K_2$ | $K_3$ | $K_4$ | $K_5$ | **m** | **q** | $\gamma$ | **Estimator** | **PRE (POPI)** | **PRE (POPII)** |
| 1 | 0 | 0 | | | | | | | | | $\bar{y}$ | 100 | 100 |
| 0 | 1 | 0 | 1 | 1 | 0 | | | 1 | | | $t_{NGR}$ | 11.13 | 8.85 |
| | | | 1 | 1 | 0 | | | -1 | | | $t_{NGP}$ | 7.48 | 12.15 |
| 0 | 0 | 1 | | | | | | | 1 | 0 | $t_{1d(1,0)}$ | 26.84 | 5.42 |
| | | | | | | | | | -1 | 0 | $t_{1d(-1,0)}$ | 23.75 | 5.87 |
| | | | | | | 1 | 0 | | 1 | 1 | $t_{2d(1,1)}$ | 82.55 | 1.23 |
| | | | | | | 1 | 0 | | 1 | -1 | $t_{2d(1,-1)}$ | 8.56 | 8.46 |
| | | | | | | 1 | 0 | | 0 | 1 | $t_{2d(0,1)}$ | 22.54 | 6.57 |
| | | | | | | 1 | 0 | | 0 | -1 | $t_{2d(0,-1)}$ | 82.56 | 7.45 |
| $w_0$ | $w_1$ | $w_2$ | 1 | 1 | 1 | 1 | 1 | 1 | 1 | 1 | $\mathbf{t_{pd}}$ **optimum** | **112.55** | **106.89** |

## 6. Conclusion

In this paper, we have proposed an unbiased estimator $t_p$ and $t_{pd}$ using information on the auxiliary attribute(s) in case of single phase and double phase sampling respectively. Expressions for bias and MSE's of the proposed estimators are derived up to first degree of approximation. From theoretical discussion and empirical study we conclude that the proposed estimators $t_p$ and $t_{pd}$ under optimum conditions perform better than other estimators considered in the article.

## Appendix A.

**Some members of the proposed family of estimators -**

**Some members (ratio-type) of the class** $t_1$

**When** $w_0 = 0,\ w_1 = 1,\ w_2 = 0\ \ \alpha = 1,$

| $K_1$ | $K_3$ | **Estimators (** $K_2 = 1$ **)** | **Estimators (** $K_2 = -1$ | **PRE'S** $K_2 = 1$ | **PRE'S** $K_2 = -1$ |
|---|---|---|---|---|---|
| 1 | $C_p$ | $t_{1a1} = \bar{y}\left[\frac{P + C_p}{p + C_p}\right]$ | $t_{1b1} = \bar{y}\left[\frac{P - C_p}{p - C_p}\right]$ | 134.99 | 72.50 |
| 1 | $\beta_2(\phi)$ | $t_{1a2} = \bar{y}\left[\frac{P + \beta_2(\phi)}{p + \beta_2(\phi)}\right]$ | $t_{1b2} = \bar{y}\left[\frac{P - \beta_2(\phi)}{p - \beta_2(\phi)}\right]$ | 111.62 | 89.34 |
| $\beta_2(\phi)$ | $C_p$ | $t_{1a3} = \bar{y}\left[\frac{P\beta_2(\phi) + C_p}{p\beta_2(\phi) + C_p}\right]$ | $t_{1b3} = \bar{y}\left[\frac{P\beta_2(\phi) - C_p}{p\beta_2(\phi) - C_p}\right]$ | 226.28 | 12.99 |
| $C_p$ | $\beta_2(\phi)$ | $t_{1a4} = \bar{y}\left[\frac{PC_p + \beta_2(\phi)}{pC_p + \beta_2(\phi)}\right]$ | $t_{1b4} = \bar{y}\left[\frac{PC_p - \beta_2(\phi)}{pC_p - \beta_2(\phi)}\right]$ | 126.66 | 77.93 |
| 1 | $\rho_{pb}$ | $t_{1a5} = \bar{y}\left[\frac{P + \rho_{pb}}{p + \rho_{pb}}\right]$ | $t_{1b5} = \bar{y}\left[\frac{P - \rho_{pb}}{p - \rho_{pb}}\right]$ | 207.46 | 39.13 |
| NP | $S_\phi$ | $t_{1a6} = \bar{y}\left[\frac{NP^2 + S_\phi}{NPp + S_\phi}\right]$ | $t_{1b6} = \bar{y}\left[\frac{NP^2 - S_\phi}{NPp - S_\phi}\right]$ | 18.14 | 6.86 |
| NP | F | $t_{1a7} = \bar{y}\left[\frac{NP^2 + f}{NPp + f}\right]$ | $t_{1b7} = \bar{y}\left[\frac{NP^2 - f}{NPp - f}\right]$ | 13.79 | 10.85 |
| $\beta_2(\phi)$ | $K_{pb}$ | $t_{1a8} = \bar{y}\left[\frac{\beta_2(\phi)P + K_{pb}}{\beta_2(\phi)p + K_{pb}}\right]$ | $t_{1b8} = \bar{y}\left[\frac{\beta_2(\phi)P - K_{pb}}{\beta_2(\phi)p - K_{pb}}\right]$ | 24.15 | 5.40 |
| NP | $K_{pb}$ | $t_{1a9} = \bar{y}\left[\frac{NP^2 + K_{pb}}{NPp + K_{pb}}\right]$ | $t_{1b9} = \bar{y}\left[\frac{NP^2 - K_{pb}}{NPp - K_{pb}}\right]$ | 18.62 | 7.78 |
| N | 1 | $t_{1a10} = \bar{y}\left[\frac{NP+1}{Np+1}\right]$ | $t_{1b10} = \bar{y}\left[\frac{NP-1}{Np-1}\right]$ | 15.93 | 9.26 |
| N | $C_p$ | $t_{1a11} = \bar{y}\left[\frac{NP+C_p}{Np+C_p}\right]$ | $t_{1b11} = \bar{y}\left[\frac{NP-C_p}{Np-C_p}\right]$ | 19.79 | 6.86 |
| N | $\rho_{pb}$ | $t_{1a12} = \bar{y}\left[\frac{NP+\rho_{pb}}{Np+\rho_{pb}}\right]$ | $t_{1b12} = \bar{y}\left[\frac{NP-\rho_{pb}}{Np-\rho_{pb}}\right]$ | 15.18 | 9.78 |
| N | $S_\phi$ | $t_{1a13} = \bar{y}\left[\frac{NP+S_\phi}{Np+S_\phi}\right]$ | $t_{1b13} = \bar{y}\left[\frac{NP-S_\phi}{Np-S_\phi}\right]$ | 12.34 | 10.96 |
| N | F | $t_{1a14} = \bar{y}\left[\frac{NP+f}{Np+f}\right]$ | $t_{1b14} = \bar{y}\left[\frac{NP-f}{Np-f}\right]$ | 12.99 | 11.54 |
| N | g=1-f | $t_{1a15} = \bar{y}\left[\frac{NP+g}{Np+g}\right]$ | $t_{1b15} = \bar{y}\left[\frac{NP-g}{Np-g}\right]$ | 15.81 | 9.34 |
| N | $K_{pb}$ | $t_{1a16} = \bar{y}\left[\frac{NP+K_{pb}}{Np+K_{pb}}\right]$ | $t_{1b16} = \bar{y}\left[\frac{NP-K_{pb}}{Np-K_{pb}}\right]$ | 13.52 | 11.10 |
| N | $\rho_{pb}$ | $t_{1a17} = \bar{y}\left[\frac{nP+\rho_{pb}}{np+\rho_{pb}}\right]$ | $t_{1b17} = \bar{y}\left[\frac{nP-\rho_{pb}}{np-\rho_{pb}}\right]$ | 25.13 | 4.86 |
| N | $S_\phi$ | $t_{1a18} = \bar{y}\left[\frac{nP+\rho_{pb}}{np+\rho_{pb}}\right]$ | $t_{1b18} = \bar{y}\left[\frac{nP-\rho_{pb}}{np-\rho_{pb}}\right]$ | 14.98 | 8.81 |
| N | F | $t_{1a19} = \bar{y}\left[\frac{nP+f}{np+f}\right]$ | $t_{1b19} = \bar{y}\left[\frac{nP-f}{np-f}\right]$ | 13.38 | 11.20 |
| N | g=1-f | $t_{1a20} = \bar{y}\left[\frac{nP+g}{np+g}\right]$ | $t_{1b20} = \bar{y}\left[\frac{nP-g}{np-g}\right]$ | 29.13 | 3.68 |
| N | $K_{pb}$ | $t_{1a21} = \bar{y}\left[\frac{nP+K_{pb}}{np+K_{pb}}\right]$ | $t_{1b21} = \bar{y}\left[\frac{nP-K_{pb}}{np-K_{pb}}\right]$ | 15.87 | 9.39 |
| $\beta_2(\phi)$ | P | $t_{1a22} = \bar{y}\left[\frac{\beta_2(\phi)P+P}{\beta_2(\phi)p+P}\right]$ | $t_{1b22} = \bar{y}\left[\frac{\beta_2(\phi)P-P}{\beta_2(\phi)p-P}\right]$ | 16.80 | 7.63 |
| NP | P | $t_{1a23} = \bar{y}\left[\frac{NP^2+P}{NPp+P}\right]$ | $t_{1b23} = \bar{y}\left[\frac{NP^2-P}{NPp-P}\right]$ | 15.93 | 9.26 |
| N | P | $t_{1a24} = \bar{y}\left[\frac{NP+P}{Np+P}\right]$ | $t_{1b24} = \bar{y}\left[\frac{NP-P}{Np-P}\right]$ | 13.23 | 11.32 |
| N | P | $t_{1a25} = \bar{y}\left[\frac{nP+P}{np+P}\right]$ | $t_{1b25} = \bar{y}\left[\frac{nP-P}{np-P}\right]$ | 14.51 | 10.28 |

**Appendix B.**

**Some members (product-type) of the class** $t_1$

**When** $w_0 = 0,\ w_1 = 1,\ w_2 = 0$

$\alpha = -1,$

| $K_1$ | $K_3$ | **Estimators** ($K_2 = 1$) | **Estimators** ($K_2 = -1$) | **PRE'S** $K_2 = 1$ | **PRE'S** $K_2 = -1$ |
|---|---|---|---|---|---|
| 1 | $C_p$ | $t_{1c1} = \bar{y}\left[\frac{p+C_p}{P+C_p}\right]$ | $t_{1d1} = \bar{y}\left[\frac{p-C_p}{P-C_p}\right]$ | 35.54 | 9.93 |
| 1 | $\beta_2(\phi)$ | $t_{1c2}=\bar{y}\left[\frac{p+\beta_2(\phi)}{P+\beta_2(\phi)}\right]$ | $t_{1d2}=\bar{y}\left[\frac{p-\beta_2(\phi)}{P-\beta_2(\phi)}\right]$ | 110.12 | 101.54 |
| $\beta_2(\phi)$ | $C_p$ | $t_{1c3}=\bar{y}\left[\frac{p\beta_2(\phi)+C_p}{P\beta_2(\phi)+C_p}\right]$ | $t_{1d3}=\bar{y}\left[\frac{p\beta_2(\phi)-C_p}{P\beta_2(\phi)-C_p}\right]$ | 6.09 | 0.127 |
| $C_p$ | $\beta_2(\phi)$ | $t_{1c4}=\bar{y}\left[\frac{pC_p+\beta_2(\phi)}{PC_p+\beta_2(\phi)}\right]$ | $t_{1d4}=\bar{y}\left[\frac{pC_p-\beta_2(\phi)}{PC_p-\beta_2(\phi)}\right]$ | 99.38 | 82.52 |
| 1 | $\rho_{pb}$ | $t_{1c5}=\bar{y}\left[\frac{p+\rho_{pb}}{P+\rho_{pb}}\right]$ | $t_{1d5}=\bar{y}\left[\frac{p-\rho_{pb}}{P-\rho_{pb}}\right]$ | 0.00135 | 5.42 |
| NP | $S_\phi$ | $t_{1c6}=\bar{y}\left[\frac{NPp+S_\phi}{NP^2+S_\phi}\right]$ | $t_{1d6}=\bar{y}\left[\frac{NPp-S_\phi}{NP^2-S_\phi}\right]$ | 2.53 | 1.23 |
| NP | f | $t_{1c7}=\bar{y}\left[\frac{NPp+f}{NP^2+f}\right]$ | $t_{1d7}=\bar{y}\left[\frac{NPp-f}{NP^2-f}\right]$ | 2.03 | 1.52 |
| $\beta_2(\phi)$ | $K_{pb}$ | $t_{1c8}=\bar{y}\left[\frac{\beta_2(\phi)p+K_{pb}}{\beta_2(\phi)P+K_{pb}}\right]$ | $t_{1d8}=\bar{y}\left[\frac{\beta_2(\phi)P-K_{pb}}{\beta_2(\phi)p-K_{pb}}\right]$ | 1.83 | 1.68 |
| NP | $K_{pb}$ | $t_{1c9}=\bar{y}\left[\frac{NPp+K_{pb}}{NP^2+K_{pb}}\right]$ | $t_{1d9}=\bar{y}\left[\frac{NPp-K_{pb}}{NP^2-K_{pb}}\right]$ | 1.89 | 1.63 |
| N | 1 | $t_{1c10}=\bar{y}\left[\frac{Np+1}{NP+1}\right]$ | $t_{1d10}=\bar{y}\left[\frac{Np-1}{NP-1}\right]$ | 2.37 | 1.30 |
| 1 | $\beta_2(\phi)$ | $t_{1c2}=\bar{y}\left[\frac{p+\beta_2(\phi)}{P+\beta_2(\phi)}\right]$ | $t_{1d2}=\bar{y}\left[\frac{p-\beta_2(\phi)}{P-\beta_2(\phi)}\right]$ | 110.12 | 101.54 |
| $\beta_2(\phi)$ | $C_p$ | $t_{1c3}=\bar{y}\left[\frac{p\beta_2(\phi)+C_p}{P\beta_2(\phi)+C_p}\right]$ | $t_{1d3}=\bar{y}\left[\frac{p\beta_2(\phi)-C_p}{P\beta_2(\phi)-C_p}\right]$ | 6.09 | 0.127 |
| $C_p$ | $\beta_2(\phi)$ | $t_{1c4}=\bar{y}\left[\frac{pC_p+\beta_2(\phi)}{PC_p+\beta_2(\phi)}\right]$ | $t_{1d4}=\bar{y}\left[\frac{pC_p-\beta_2(\phi)}{PC_p-\beta_2(\phi)}\right]$ | 99.38 | 82.52 |
| N | $C_p$ | $t_{1c11}=\bar{y}\left[\frac{Np+C_p}{NP+C_p}\right]$ | $t_{1d11}=\bar{y}\left[\frac{Np-C_p}{NP-C_p}\right]$ | 2.50 | 1.23 |
| N | $\rho_{pb}$ | $t_{1c12}=\bar{y}\left[\frac{Np+\rho_{pb}}{NP+\rho_{pb}}\right]$ | $t_{1d12}=\bar{y}\left[\frac{Np-\rho_{pb}}{NP-\rho_{pb}}\right]$ | 1.79 | 1.70 |
| N | $S_\phi$ | $t_{1c13}=\bar{y}\left[\frac{Np+S_\phi}{NP+S_\phi}\right]$ | $t_{1d13}=\bar{y}\left[\frac{Np-S_\phi}{NP-S_\phi}\right]$ | 2.16 | 1.44 |
| N | f | $t_{1c14}=\bar{y}\left[\frac{Np+f}{NP+f}\right]$ | $t_{1d14}=\bar{y}\left[\frac{Np-f}{NP-f}\right]$ | 1.98 | 1.56 |
| N | g=1-f | $t_{1c15}=\bar{y}\left[\frac{Np+g}{NP+g}\right]$ | $t_{1d15}=\bar{y}\left[\frac{Np-g}{NP-g}\right]$ | 2.34 | 1.32 |
| N | $K_{pb}$ | $t_{1c16}=\bar{y}\left[\frac{Np+K_{pb}}{NP+K_{pb}}\right]$ | $t_{1d16}=\bar{y}\left[\frac{Np-K_{pb}}{NP-K_{pb}}\right]$ | 1.93 | 1.60 |
| N | $\rho_{pb}$ | $t_{1c17}=\bar{y}\left[\frac{np+\rho_{pb}}{nP+\rho_{pb}}\right]$ | $t_{1d17}=\bar{y}\left[\frac{np-\rho_{pb}}{nP-\rho_{pb}}\right]$ | 1.49 | 1.96 |
| N | $S_\phi$ | $t_{1c18}=\bar{y}\left[\frac{np+\rho_{pb}}{nP+\rho_{pb}}\right]$ | $t_{1d18}=\bar{y}\left[\frac{np-\rho_{pb}}{nP-\rho_{pb}}\right]$ | 2.65 | 1.14 |
| N | f | $t_{1c19}=\bar{y}\left[\frac{np+f}{nP+f}\right]$ | $t_{1d19}=\bar{y}\left[\frac{np-f}{nP-f}\right]$ | 2.06 | 1.51 |
| N | g=1-f | $t_{1c20}=\bar{y}\left[\frac{np+g}{nP+g}\right]$ | $t_{1d20}=\bar{y}\left[\frac{np-g}{nP-g}\right]$ | 3.29 | 0.84 |
| N | $K_{pb}$ | $t_{1c21}=\bar{y}\left[\frac{np+K_{pb}}{nP+K_{pb}}\right]$ | $t_{1d21}=\bar{y}\left[\frac{np-K_{pb}}{nP-K_{pb}}\right]$ | 1.88 | 1.63 |
| $\beta_2(\phi)$ | P | $t_{1c22} = \bar{y}\left[\frac{\beta_2(\phi)p + P}{\beta_2(\phi)P + P}\right]$ | $t_{1d22} = \bar{y}\left[\frac{\beta_2(\phi)p - P}{\beta_2(\phi)P - P}\right]$ | 2.99 | 0.97 |
| NP | P | $t_{1c23} = \bar{y}\left[\frac{NPp + P}{NP^2 + P}\right]$ | $t_{1d23} = \bar{y}\left[\frac{NPp - P}{NP^2 - P}\right]$ | 2.37 | 1.30 |
| N | P | $t_{1c24} = \bar{y}\left[\frac{Np + P}{NP + P}\right]$ | $t_{1d24} = \bar{y}\left[\frac{Np - P}{NP - P}\right]$ | 2.11 | 1.47 |
| N | P | $t_{1c25} = \bar{y}\left[\frac{np + P}{nP + P}\right]$ | $t_{1d25} = \bar{y}\left[\frac{np - P}{nP - P}\right]$ | 2.49 | 1.23 |

**Appendix C.**

**Some members (product-type) of the class $t_2$**

**When** $w_0 = 0,\ w_1 = 0,\ w_2 = 1$

| $K_4$ | $K_5$ | **Estimators** $(\beta = 1, \lambda = -1)$ | PRE'S |
|---|---|---|---|
| 1 | $C_p$ | $t_{21} = \bar{y}\left(2 - \frac{p}{P}\exp\left[\frac{p - P}{p + P + 2C_p}\right]\right)$ | 12.42 |
| 1 | $\beta_2(\phi)$ | $t_{22} = \bar{y}\left(2 - \frac{p}{P}\exp\left[\frac{p - P}{p + P + 2\beta_2(\phi)}\right]\right)$ | 11.92 |
| $\beta_2(\phi)$ | $C_p$ | $t_{23} = \bar{y}\left(2 - \frac{p}{P}\exp\left[\frac{\beta_2(\phi)(p - P)}{\beta_2(\phi)(p + P) + 2C_p}\right]\right)$ | 16.29 |
| $C_p$ | $\beta_2(\phi)$ | $t_{24} = \bar{y}\left(2 - \frac{p}{P}\exp\left[\frac{C_p(p - P)}{C_p(p + P) + 2\beta_2(\phi)}\right]\right)$ | 12.53 |
| 1 | $\rho_{pb}$ | $t_{25} = \bar{y}\left(2 - \frac{p}{P}\exp\left[\frac{(p-P)}{(p+P)+2\rho_{pb}}\right]\right)$ | 13.86 |
| NP | $S_\phi$ | $t_{26} = \bar{y}\left(2 - \frac{p}{P}\exp\left[\frac{NP(p-P)}{NP(p+P)+2S_\phi}\right]\right)$ | 44.46 |
| NP | F | $t_{27} = \bar{y}\left(2 - \frac{p}{P}\exp\left[\frac{NP(p-P)}{NP(p+P)+2f}\right]\right)$ | 61.84 |
| $\beta_2(\phi)$ | $K_{pb}$ | $t_{28} = \bar{y}\left(2 - \frac{p}{P}\exp\left[\frac{\beta_2(\phi)(p-P)}{\beta_2(\phi)(p+P)+2K_{pb}}\right]\right)$ | 40.17 |
| NP | $K_{pb}$ | $t_{29} = \bar{y}\left(2 - \frac{p}{P}\exp\left[\frac{NP(p-P)}{NP(p+P)+2K_{pb}}\right]\right)$ | 48.09 |
| N | 1 | $t_{210} = \bar{y}\left(2 - \frac{p}{P}\exp\left[\frac{N(p-P)}{N(p+P)+2}\right]\right)$ | 54.10 |
| N | $C_p$ | $t_{211} = \bar{y}\left(2 - \frac{p}{P}\exp\left[\frac{N(p-P)}{N(p+P)+2C_p}\right]\right)$ | 44.84 |
| N | $\rho_{pb}$ | $t_{212} = \bar{y}\left(2 - \frac{p}{P}\exp\left[\frac{N(p-P)}{N(p+P)+2\rho_{pb}}\right]\right)$ | 56.48 |
| N | $S_\phi$ | $t_{213} = \bar{y}\left(2 - \frac{p}{P}\exp\left[\frac{N(p-P)}{N(p+P)+2S_\phi}\right]\right)$ | 62.40 |
| N | F | $t_{214} = \bar{y}\left(2 - \frac{p}{P}\exp\left[\frac{N(p-P)}{N(p+P)+2f}\right]\right)$ | 65.67 |
| N | g=1-f | $t_{215}=\bar{y}\left(2-\frac{p}{P}\exp\left[\frac{N(p-P)}{N(p+P)+2g}\right]\right)$ | 54.47 |
| N | $K_{pb}$ | $t_{216}=\bar{y}\left(2-\frac{p}{P}\exp\left[\frac{N(p-P)}{N(p+P)+2K_{pb}}\right]\right)$ | 63.21 |
| N | $\rho_{pb}$ | $t_{217}=\bar{y}\left(2-\frac{p}{P}\exp\left[\frac{n(p-P)}{n(p+P)+2\rho_{pb}}\right]\right)$ | 38.59 |
| N | $S_{\phi}$ | $t_{218}=\bar{y}\left(2-\frac{p}{P}\exp\left[\frac{n(p-P)}{n(p+P)+2S_{\phi}}\right]\right)$ | 52.19 |
| N | F | $t_{219}=\bar{y}\left(2-\frac{p}{P}\exp\left[\frac{n(p-P)}{n(p+P)+2f}\right]\right)$ | 63.74 |
| N | g=1-f | $t_{220}=\bar{y}\left(2-\frac{p}{P}\exp\left[\frac{n(p-P)}{n(p+P)+2g}\right]\right)$ | 35.33 |
| N | $K_{pb}$ | $t_{221}=\bar{y}\left(2-\frac{p}{P}\exp\left[\frac{n(p-P)}{n(p+P)+2K_{pb}}\right]\right)$ | 54.68 |
| $\beta_2(\phi)$ | P | $t_{222}=\bar{y}\left(2-\frac{p}{P}\exp\left[\frac{\beta_2(\phi)(p-P)}{\beta_2(\phi)(p+P)+2P}\right]\right)$ | 47.53 |
| NP | P | $t_{223}=\bar{y}\left(2-\frac{p}{P}\exp\left[\frac{NP(p-P)}{NP(p+P)+2P}\right]\right)$ | 54.10 |
| N | P | $t_{224}=\bar{y}\left(2-\frac{p}{P}\exp\left[\frac{N(p-P)}{N(p+P)+2P}\right]\right)$ | 64.43 |
| N | P | $t_{225}=\bar{y}\left(2-\frac{p}{P}\exp\left[\frac{n(p-P)}{n(p+P)+2P}\right]\right)$ | 58.91 |

In addition to above estimators a large number of estimators can also be generated from the proposed estimators just by putting different values of constants $w_i$'s, $h_i$'s $K_1, K_2, K_3, K_4, K_5, \alpha, \beta$ and $\lambda$.